\documentclass[a4paper,11pt]{article}
\usepackage{pos}
\usepackage{lineno}
\usepackage{enumitem}
\usepackage{wrapfig}
\usepackage{siunitx}
\usepackage{float}

\title{Characterization of downward Terrestrial Gamma-ray Flashes detected at the Pierre Auger Observatory}
 \ShortTitle{Downward TGF at the Pierre Auger Observatory}

\author[]{Roberta Colalillo$^{*,a,b}$ for the Pierre Auger Collaboration$^c$}
\author[d]{Joseph Dwyer}
\author[e]{John Ortberg}
\author[e]{David M Smith}

\affiliation[a]{Università degli Studi di Napoli “Federico II”, Dipartimento di Fisica “E. Pancini”, \\Via Cintia, 80126 Napoli, Italy}
\affiliation[b]{INFN, Sezione di Napoli, Via Cintia, 80126 Napoli, Italy}
\affiliation[c]{Observatorio Pierre Auger, Av.\ San Mart{\'\i}n Norte 304, 5613 Malarg\"ue, Argentina\\
Full author list: {\rm\url{https://www.auger.org/archive/authors_icrc_2025.html}}}
\affiliation[d]{Department of Physics \& Astronomy and Space Science Center (EOS), University of New Hampshire, \\8 College Road, 03824 Durham, NH, USA}
\affiliation[e]{Physics Department and Santa Cruz Institute for Particle Physics, University of California, \\ 1156 High Street, Santa Cruz, CA 95064, USA}

\emailAdd{spokespersons@auger.org}

\abstract{Downward Terrestrial Gamma-ray Flashes (TGFs) are sub-millisecond bursts of MeV gamma rays produced in thunderclouds. According to the Relativistic Runaway Electron Avalanche model, gamma rays are produced, via bremsstrahlung, from electron cascades activated by a relativistic “seed” electron. It is not clear what mechanism is responsible for the acceleration of electrons to relativistic energies in electric discharges. To better understand the acceleration sites and the TGF production mechanisms, it is critically important to identify the TGF source position and geometry in the atmosphere and to study the gamma emission characteristics. The Surface Detector of the Pierre Auger Observatory, with its 1600 water-Cherenkov detectors very sensitive to high-energy photons and with a very fine time-sampling, is a valuable instrument to study downward TGFs. The possibility to analyze the radiation emission in detail led to the observation of the first TGFs with an asymmetric azimuthal structure, suggesting a complex source different from the initially hypothesized downward beam. We report on these observations and the new perspectives which may open with the incorporation of new instruments at the Auger site to study lightning development alongside gamma emission, and the increasingly detailed data provided by satellites and global lightning networks.}

\ConferenceLogo{PoS_ICRC2025_logo}

\FullConference{%
39th International Cosmic Ray Conference (ICRC2025)\\
15 -- 24 July, 2025\\
Geneva, Switzerland}

\begin{document}
\maketitle
\section{The observation of downward TGFs at the Pierre Auger Observatory}
Terrestrial Gamma-ray Flashes (TGFs) are intense bursts of gamma radiation originating in Earth's atmosphere, typically associated with thunderstorms and lightning activity. They can last from 10's of $\mu s$ to a few milliseconds. They can be directed upwards and downwards. Upward TGFs were discovered in 1994 by satellite-based instruments~\cite{BATSE} and extensively studied. Downward TGFs were observed in 2004 for the first time and their statistics are still low. TGFs provide critical insights into high-energy processes occurring within thunderstorms, offering a unique window into the complex interactions between electrical phenomena and high-energy particles. The development of a TGF is explained by the RREA (Relativistic Runaway Electron Avalanche) mechanism. But how is the electron cascade triggered? At what stage of a lightning development does this occur? Why are TGFs so bright? Additionally, why are some lightning of the same type accompanied by TGFs and others not? These questions remain not fully understood.\\
Observing TGFs from the ground is essential for complementing satellite data, enabling detailed studies of their spatial and temporal characteristics, and improving our understanding of their production mechanisms. 
The Pierre Auger Observatory~\cite{PAO}, primarily designed for ultra-high-energy cosmic ray detection, has proven to be an excellent instrument for detecting TGFs. Its extensive Surface Detector (SD) array and atmospheric monitoring systems~\cite{atmoHEAD24_moni} present a valuable opportunity to detect and analyze high-energy atmospheric events, contributing significantly to the growing field of TGF research. From space, even if a TGF was observed by 2 or 3 satellites positioned hundreds of kilometers apart, its source would be reconstructed with a large uncertainty. Ground observatories such as the Pierre Auger Observatory, on the other hand, allow for detailed sampling of the TGF emission, enabling the study of the signal distribution and precise reconstruction of the TGF source position. Additionally, water-Cherenkov Detectors (WCDs), which are the fundamental units of the Auger SD, are highly sensitive to gamma rays with energies above 1 MeV —such as those produced by TGFs— and can detect signals with a temporal resolution of 10's of ns.\\
Downward TGFs detected at the Pierre Auger Observatory exhibit a signature and temporal evolution that are remarkably different from those of a shower generated by an ultra-high-energy cosmic ray~\cite{ICRC2021}. The number of triggered SD stations is significantly higher than that typical of extensive air showers, with the footprint spanning approximately 200 $km^2$. Additionally, the signals observed in the WCDs last over 10 $\mu s$, which is an order of magnitude longer than the duration of signals produced by cosmic-ray showers. For this reason, these stations are called "long-signal stations". Finally, these events are characterized by the presence of at least one "lightning station" with a signal dominated by high-frequency noise. For each station, we have three signals from three different PMTs looking into the water from above. They collect the Cherenkov light produced by charged particles crossing the station. 
The water depth of 1.2 m allows high-energy photons to convert into electron-positron pairs with high probability and to produce a signal. The signals are then digitized by a 40 MHz FADC. The SD trigger system employs a hierarchical logic~\cite{SDtrigger}: two local station triggers (T1 and T2) identify candidate signals based on signal amplitude and timing, which are then combined into a global T3 trigger to confirm the presence of a genuine event (cosmic-ray shower or TGF) across multiple detectors. The event formation, the communication with the stations, and the saving of all information related to an event are managed by the CDAS (Central Data Acquisition System).\\
The installation of AugerPrime, the major upgrade of the Pierre Auger Observatory, has recently been completed. It includes the addition of new detector components, scintillators and radio antennas, and the replacement of the electronic boards in each SD station. Within AugerPrime, there is also the intention to continue detecting downward TGFs by leveraging the potential offered by the new electronics and utilizing the new detectors, particularly the scintillators, to compare our signals with those from other experiments that detect downward TGFs using this technology.

\section{Characterization of TGF footprint and theoretical implications}
\label{sec:TGFinterpr}
\begin{figure*}[!h]
\centering \includegraphics[width=1.0
\columnwidth]{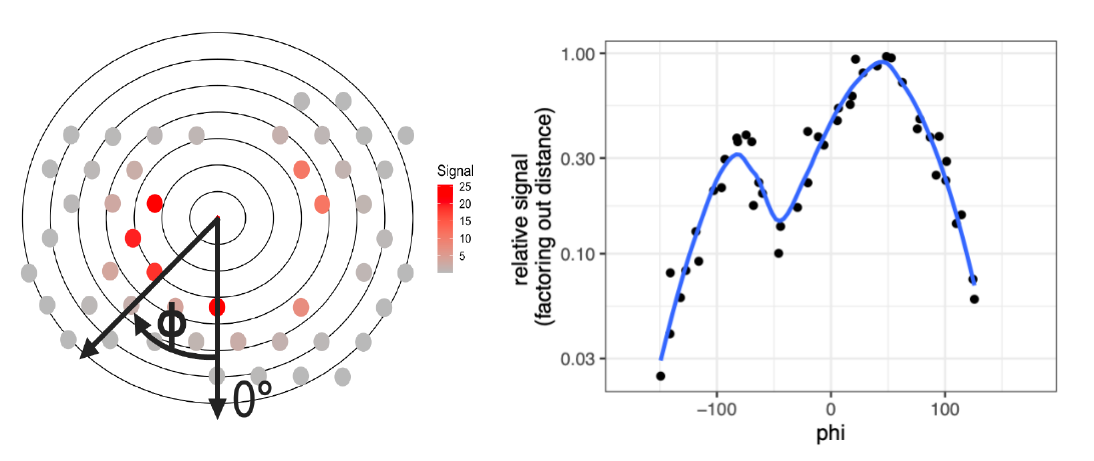}
\caption{The left panel shows a TGF footprint. The dots represents the active stations and their color is proportional to the signal. It is clear that the signal distribution is asymmetric. At the same distance from the center is not the same at different azimuthal angles. Moreover, the signal variation as function of the azimuthal angle presents two peaks as shown on the right panel.}
\label{AsymTGF}
\end{figure*}

We have observed less than 2 events/year, while at least 30 events/year are expected considering the known lightning rate at the Observatory and a lightning/TGFs ratio of approximately $10^3$. Moreover, most of our TGF events show a lack of signals in the center of the footprint. Both are due to electronics, trigger, data acquisition, or post-acquisition processing of the Auger SD which is optimized for rate, shape and signals of cosmic-ray showers~\cite{ICRC2021}. Despite an inefficient acquisition for TGF events, further complicated by a modification in the SD trigger in 2014, the 22 events collected before 2017 allowed us to put constraints on the current understanding of downward TGFs. The rising and falling edge of our TGF long signals can be fit with an exponential function and this is compatible with the relativistic feedback mechanism, one of the models that try to explain the TGF origin~\cite{ICRC2023}. Moreover, using instruments available at the Auger Observatory for monitoring the atmosphere, we verified that, at the time of our events, there were thunderclouds with a base at 1-2 km from the ground. This is consistent with the source height we reconstruct from the arrival times of TGF signals at our stations, and it aligns with what is expected for downward TGFs. We also studied Convective Available Potential Energy (CAPE) and the Precipitable Water Vapor (PWV) in coincidence with Auger events and we observed that they occur at the beginning of a thunderstorm. At the moment, this result might be biased by the described acquisition issues, which can cause us to detect only a portion of the events produced by a certain thunderstorm system. Finally, we compared the deposited energy at the ground of the Auger TGF events as a function of the distance from the source - the center of the footprint - with the deposited energy obtained simulating a standard TGF ($10^{17}$ - $10^{18}$ photons) having a normal, wide or isotropic emission cone. The Auger TGF brightness is similar to expectations but the trend as a function of the distance from the source is not well reproduced by the standard models. Further studies were conducted to understand if a tilted TGF could explain the observed trend and first results have shown that Auger TGFs are asymmetric about their vertical axis. They also present two peaks in the azimuthal direction (see figure \ref{AsymTGF}).
This azimuthal substructure was predicted by several model but never previously observed~\cite{Dwyer21,Stadnichuk21}. 
\section{Performance of the long-signal algorithm for identification of TGF traces}
\begin{figure*}[!h]
\centering \includegraphics[width=1.0\columnwidth]{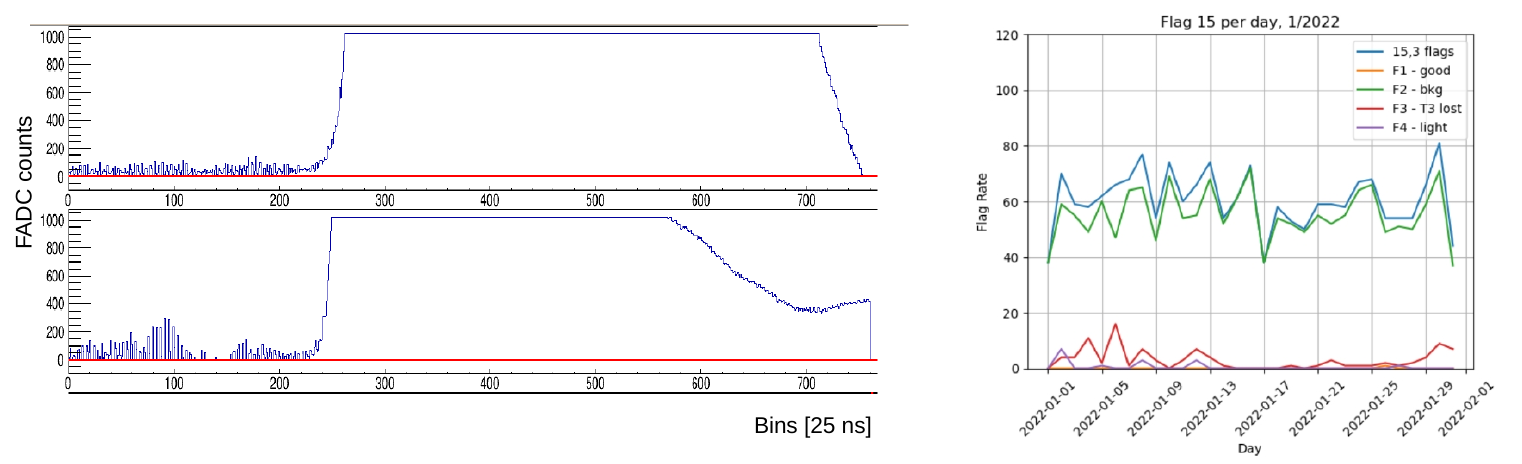}
\caption{Left: A "long signal" detected in January 2022 with the "tagger" algorithm designed to identify TGF traces and prioritize TGF acquisition in CDAS. Right: The daily "long flag" rate in January 2022.}
\label{NewLongSignal}
\end{figure*}

\begin{figure*}[!h]
\centering \includegraphics[width=1.0\columnwidth]{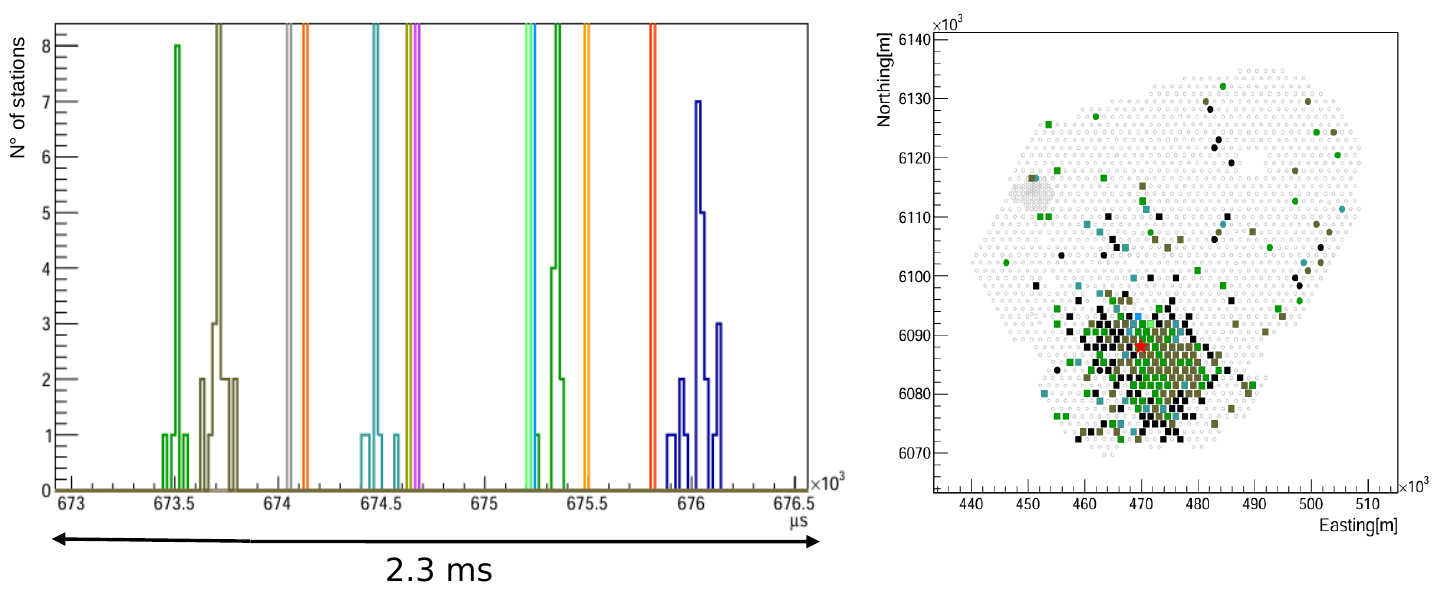}
\caption{Left: T3 burst detected in December 2021. Each histogram shows the time distribution of the stations with signal. If all event stations are "T3 lost", the event is identified by a line corresponding to the GPS time of the event. Right: The burst time scale is consistent with the time evolution of lightning and each event was detected in the same zone of the array, coincident with the position (red star) of a negative cloud to ground lightning collected by ENTLN.}
\label{BurstNeg}
\end{figure*}

\begin{figure*}[!h]
\centering \includegraphics[width=1.0\columnwidth]{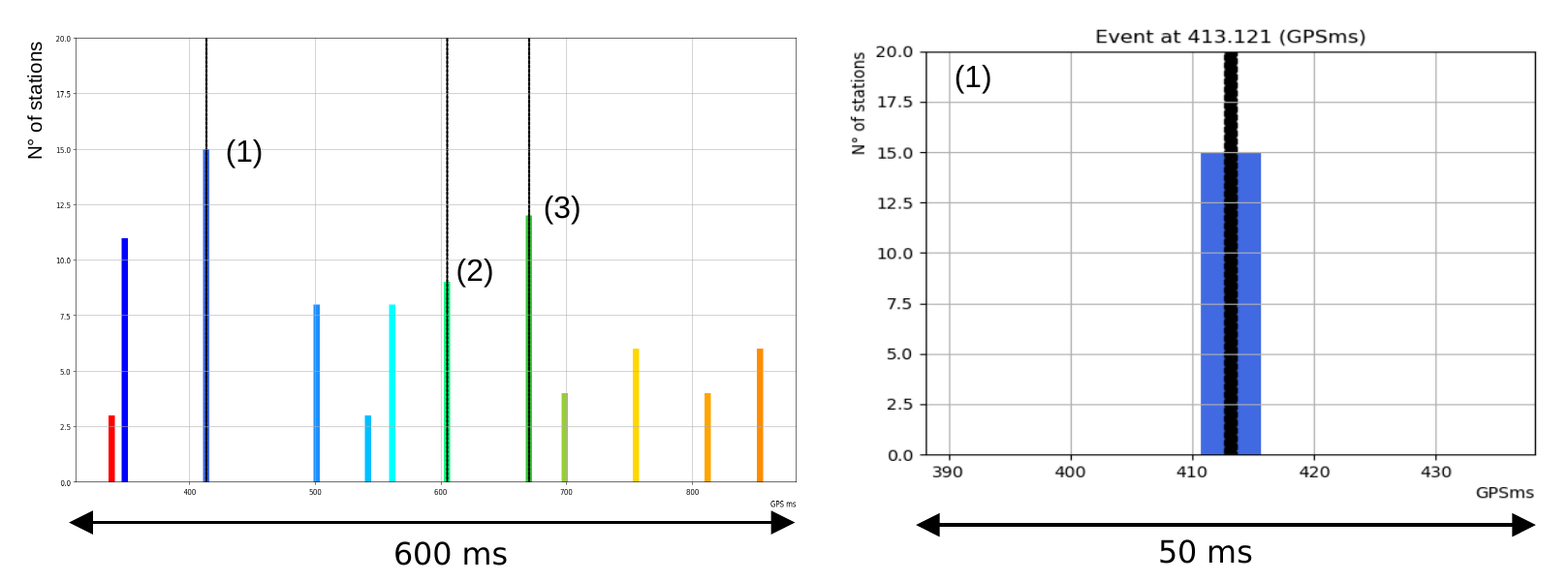}
\caption{Left: Burst of 12 events detected in February 2022. The overlapping black lines on top of the colored lines represent the timing of lightning strikes observed in conjunction with our events marked with the numbers (1), (2), (3). Right: Analyzing a shorter time window ($\pm$25 ms) around event (1) coinciding with a positive polarity lightning, we do not find any additional events as expected. We found a similar behavior for events (2) and (3).}
\label{BurstPos}
\end{figure*}

As explained in section \ref{sec:TGFinterpr}, the Auger trigger and acquisition system are not optimal for TGF detection. During a thunderstorm, the T3 rate can be up to 10 times higher than under normal conditions, and the CDAS does not have a large enough buffer to acquire all the events. It is highly likely that TGF events are not being recorded since the majority of triggers are caused by noise induced by lightning. We could increase the data-taking efficiency for TGFs by enforcing the read-out ordering and prioritizing events containing stations with long signals, which are typical of TGFs. Following this idea, an algorithm to identify long signals was implemented in WCD electronic boards in November 2021~\cite{atmoHEAD22}. The lightning noise, that causes the increase in trigger rate during thunderstorms, is a high-frequency bipolar noise that can be cancelled if a sufficiently long time interval of the trace is integrated. Therefore, in this algorithm, three large integration windows are used. The trigger condition is fulfilled if the sum of the differences between the integrals over the three windows exceeds a certain threshold. The choice of intervals, from studying in detail the available TGF sample, covers the beginning and end of the long signals and should exclude cosmic-ray signals. This algorithm is not a trigger, but a “tagger” because it uses the data recorded with the already existing triggers and just adds a new "long flag" to the existing T2s. To study the algorithm performances, we analyzed the data collected from December 2021 to March 2022. We chose to focus on this period first because these months are usually characterized by an abundant presence of lightning, and also because the subsequent lightning season was already significantly affected by the installation of AugerPrime, involving the replacement of the WCD electronics boards. We need to analyze two different types of data: the SD data cointaining all the events which passed the T3 trigger level and the T2 dumps, which store all T2 messages received at CDAS with information on the station, trigger type, and GPS microsecond, but not the signal trace. Therefore, we search for "long flags" in T2 dumps and we verify the real trace type in the corresponding SD data. In January, we found the first "new" long signals after 8 years. It was detected by two PMTs of the station and is shown on the left of fig. \ref{NewLongSignal}. 
Unfortunately, for the other stations of this event which were tagged as "long signal", we lost the trace in the T3 event. It can happen that if we have a GPS block associated to a T3 trigger, but the station nanosecond doesn't match with the trigger time to within 3 $\mu s$, the trace is not saved in T3 event and the station is tagged as "T3 lost". This often occurs for stations of "TGF candidates" due to their different temporal evolution compared to extensive air showers. On the right panel of fig. \ref{NewLongSignal}, the daily "long flag" rate is shown throughout the entire month of January. The rate is dominated by stations not having a genuine long signal, but typical signals produced by showers induced by the most energetic cosmic rays. Immediately following, there are the "T3 lost" stations, and then the "lightning stations". This result, very similar to those observed in the other analyzed months, along with the detection of a genuine long signal, leads us to conclude that the algorithm is indeed capable of selecting "long stations" while discarding "lightning stations", consistent with our initial purpose. In the future, the algorithm could be further optimized to improve the discrimination between long signals from TGFs and signals from highly energetic cosmic rays. Additionally, the rate of "T3 lost" stations should be reduced.\\
The information about the "long flag" in T2 dumps, despite the traces being lost in SD data, allowed us to select bursts of events that, in coincidence with lightning data provided by global lightning networks such as WWLLN~\cite{WWLLN} and ENTLN~\cite{ENTLN}, and the geostationary lightning mapper GLM~\cite{GLM}, enabled us once again to understand the importance of having a dedicated trigger for the detection of TGF events at Auger. A first burst was identified on December 8, 2021 (figure \ref{BurstNeg} - left: time evolution, right: footprint of the burst events). 
It is very similar to the previously detected bursts: a series of T3 events detected within a few milliseconds, with the last event being a TGF~\cite{ICRC2021}. A negative cloud to ground lightning was detected at the same GPS second of our burst and the lightning position, indicated by the red star on the right panel of figure \ref{BurstNeg}, approximately coincides with the center of the last burst event's footprint. Furthermore, by comparing the burst of events with the radio waveform detected by the ENTLN antenna closest to the Auger Observatory, we observed that the last event, the TGF as suggested by Auger observations prior to 2017, coincides with a small peak in the return stroke waveform, called "reflection pulse," which is the moment in lightning development when certain types of TGFs are expected~\cite{SmithAGU2024}. This observation is consistent with other observations by THOR experiment in Japan~\cite{THOR,similTHOR}. We also searched for a burst prior to the event containing the genuine long signal, but no evidence was found. The TGF could be associated with lightning of positive polarity for which no event bursts are expected. Finally, thanks to the "long flag", we identified a burst of 12 events on a time scale of 600 ms shown in the left panel of figure~\ref{BurstPos}.
As usual, a temporal coincidence corresponds to a spatial coincidence. However, the 600 ms timescale is not compatible with the development of a lightning strike. In fact, we found three lightning strikes detected at the times of three of our events. If we analyze a smaller time interval around the three events with a coincident lightning, we observe that each of them is isolated (right panel of fig.~\ref{BurstPos}) as the event with the genuine long signal. Further studies are necessary to better understand if we are observing one or more types of TGFs.
In March 2022, we haven't found interesting events, but the end of the Argentinian lightning season is approaching and this is confirmed by the global lightning network databases.\\
This study has shown that the Pierre Auger Observatory with an optimized acquisition system and a "tagger" for TGF long signals can strongly contribute to the classification efforts of these events which the atmospheric electricity community has actively been pursuing over these years. Additionally, in section \ref{sec:TGFinterpr} we highlighted the uniqueness of an observatory like Auger for understanding the TGF production models.\\
The rate of T3 events with at least one station tagged by the "long flag" is very low, 1-2 events per minute. In the four analyzed months, a rate of 12 events per minute was reached only once. The implementation of the "flag long" on the old WCD electronic boards has not negatively affected the detection of cosmic rays in any way. The "flag long" will be implemented on the new SD station electronic boards before the next lightning season.
\section{Future Lightning Local Network at the Auger Observatory}
TGFs and lightning emit radio waves. The TGF gamma signal is observable within a few km, radio waves over hundreds of km. This has allowed us so far to leverage information from the worldwide lightning networks. Lightning emit radio signals during various stages of the discharge. It is very important to correlate the gamma measurements with radio measurements to better understand the TGF production mechanisms. Radio emissions cover several wavelength bands and several instruments can be used for lightning studies. E-field mills, fast and slow antennas measure DC and LF signals, while interferometers and lightning mapping arrays are used in the HF/VHF band. The VLF/LF range provides different types of signals, such as energetic in-cloud pulses (EIPs) and slow pulses,  which could be related to direct emissions from TGF currents. The VHF range is very good for mapping lightning in 2 and 3D and are sensitive to fast and short range discharges. Previous studies have shown that intense VHF emissions are closely correlated with high-energy bursts such as X-rays and TGFs, particularly during negative stepped leader development~\cite{Urbani2021-yj,Wada2025}.\\
At the Pierre Auger Observatory, to improve the correlation studies between TGFs and lightning, we can already take advantage of an E-mill array~\cite{atmoHEAD24_moni}, an interferometric array of VHF antennas that will be described in detail in section \ref{sec:bolt}, and a LF/VLF instrument array will be deployed very soon~\cite{JamesAGU2024} to complete the observatory's lightning RF capabilities.
\subsection{BOLT General Information}
\label{sec:bolt}
\begin{figure}[!h]
    \centering
    \includegraphics[height = 4.53cm]{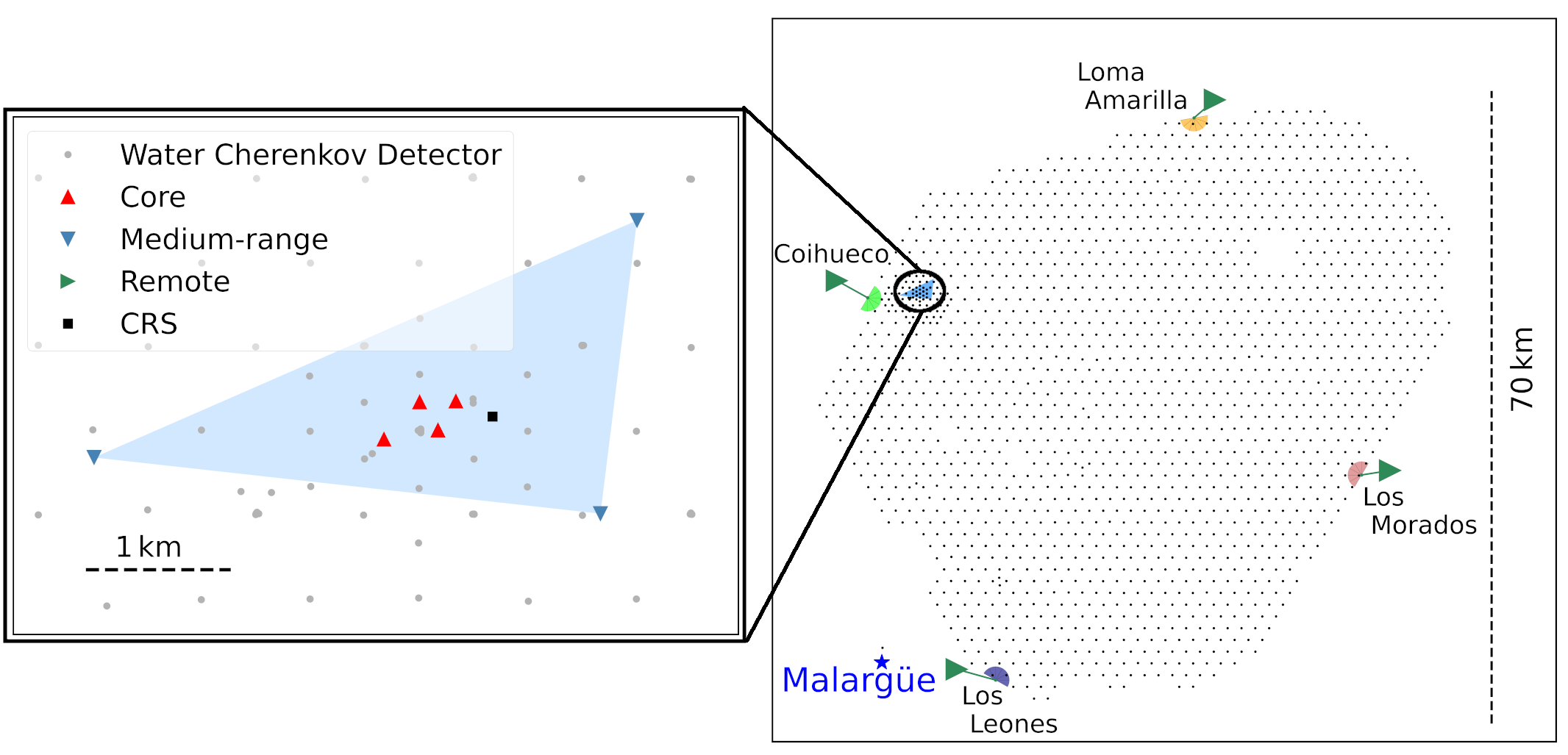}
    \hspace{.2cm}
    \includegraphics[height = 4.45cm]{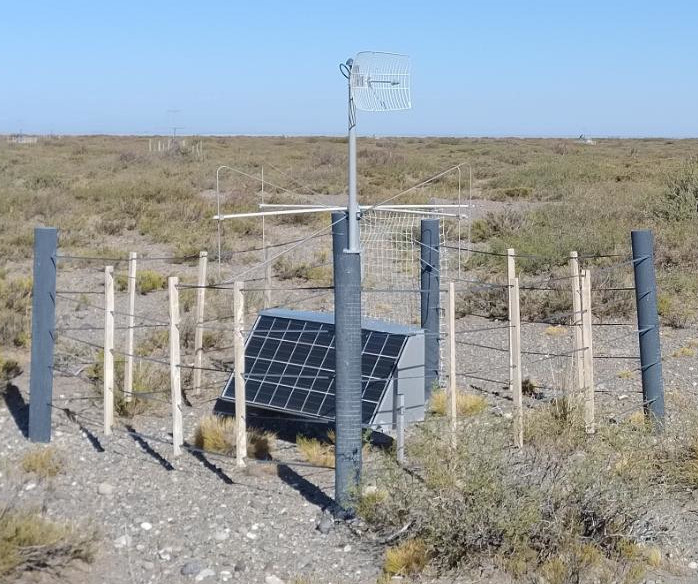}
    \caption{Left: The planned configuration of BOLT. The color-coded triangles indicate the three station groups: core (red), mid-range (blue), and remote (green). The mid-range deployment area is shaded in light orange. Light-grey circles mark the positions of water-Cherenkov surface detectors. Right: Photograph of one of the core BOLT stations deployed in the field.}
    \label{fig:BOLTmap}
\end{figure}

The \textbf{B}roadband \textbf{O}bservatory of \textbf{L}ightning and \textbf{T}GFs (BOLT) is a novel instrument under development at the Pierre Auger Observatory. 
It repurposes the existing AERA infrastructure— very high frequency (VHF) antennas operating in the 30–80\,MHz band—into a dedicated interferometric array optimised for lightning mapping, from the earliest breakdown processes to return strokes.
BOLT comprises 11 stations—four core, three mid-range, and four remote—spanning baselines from tens of metres to several tens of kilometres~\cite{atmoHEAD24}. As shown in figure~\ref{fig:BOLTmap}, all stations are co-located with SD. This geometry balances spatial resolution and coverage, enabling 4D imaging even in dense VHF source conditions expected in lightning with downward TGFs. A reference beacon ensures sub-nanosecond timing~\cite{aera_beaconpaper}, supporting metre-scale spatial resolution.
Each station includes low- and high-gain channels for a wide dynamic range, capturing both strong and weak VHF pulses without saturation. Modified AERA digitizers now record multi-second traces, enabling full lightning-flash capture and positioning BOLT to study potential TGF sources. The system is being upgraded to accept external triggers from Lightning Detection Stations~\cite{LDS_2015} and the Surface Detector.
Three core BOLT stations are operational; the rest will complete the 11-station array. Commissioning confirms stable power, communication, and synchronisation. Test readouts verify uncurtailed, multi-second waveform capture, with multi-station acquisition and interferometric reconstruction in development. Full integration will enable coordinated campaigns on lightning development and downward TGFs.

\clearpage
\section*{The Pierre Auger Collaboration}

{\footnotesize\setlength{\baselineskip}{10pt}
\noindent
\begin{wrapfigure}[11]{l}{0.12\linewidth}
\vspace{-4pt}
\includegraphics[width=0.98\linewidth]{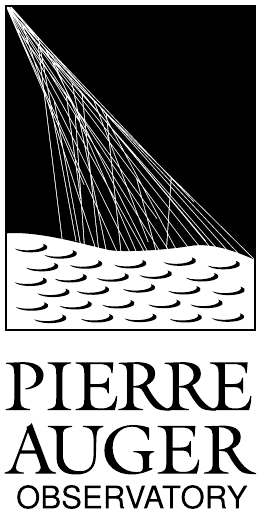}
\end{wrapfigure}
\begin{sloppypar}\noindent
A.~Abdul Halim$^{13}$,
P.~Abreu$^{70}$,
M.~Aglietta$^{53,51}$,
I.~Allekotte$^{1}$,
K.~Almeida Cheminant$^{78,77}$,
A.~Almela$^{7,12}$,
R.~Aloisio$^{44,45}$,
J.~Alvarez-Mu\~niz$^{76}$,
A.~Ambrosone$^{44}$,
J.~Ammerman Yebra$^{76}$
% ...

\end{sloppypar}
\begin{center}
\end{center}

\vspace{1ex}
\begin{description}[labelsep=0.2em,align=right,labelwidth=0.7em,labelindent=0em,leftmargin=2em,noitemsep,before={\renewcommand\makelabel[1]{##1 }}]
\item[$^{1}$] Centro At\'omico Bariloche and Instituto Balseiro (CNEA-UNCuyo-CONICET), San Carlos de Bariloche, Argentina
\item[$^{2}$] Departamento de F\'\i{}sica and Departamento de Ciencias de la Atm\'osfera y los Oc\'eanos, FCEyN, Universidad de Buenos Aires and CONICET, Buenos Aires, Argentina
\item[$^{3}$] IFLP, Universidad Nacional de La Plata and CONICET, La Plata, Argentina
\item[$^{4}$] Instituto de Astronom\'\i{}a y F\'\i{}sica del Espacio (IAFE, CONICET-UBA), Buenos Aires, Argentina
\item[$^{5}$] Instituto de F\'\i{}sica de Rosario (IFIR) -- CONICET/U.N.R.\ and Facultad de Ciencias Bioqu\'\i{}micas y Farmac\'euticas U.N.R., Rosario, Argentina
\item[$^{6}$] Instituto de Tecnolog\'\i{}as en Detecci\'on y Astropart\'\i{}culas (CNEA, CONICET, UNSAM), and Universidad Tecnol\'ogica Nacional -- Facultad Regional Mendoza (CONICET/CNEA), Mendoza, Argentina
\item[$^{7}$] Instituto de Tecnolog\'\i{}as en Detecci\'on y Astropart\'\i{}culas (CNEA, CONICET, UNSAM), Buenos Aires, Argentina
\item[$^{8}$] International Center of Advanced Studies and Instituto de Ciencias F\'\i{}sicas, ECyT-UNSAM and CONICET, Campus Miguelete -- San Mart\'\i{}n, Buenos Aires, Argentina
\item[$^{9}$] Laboratorio Atm\'osfera -- Departamento de Investigaciones en L\'aseres y sus Aplicaciones -- UNIDEF (CITEDEF-CONICET), Argentina
\item[$^{10}$] Observatorio Pierre Auger, Malarg\"ue, Argentina
\item[$^{11}$] Observatorio Pierre Auger and Comisi\'on Nacional de Energ\'\i{}a At\'omica, Malarg\"ue, Argentina
\item[$^{12}$] Universidad Tecnol\'ogica Nacional -- Facultad Regional Buenos Aires, Buenos Aires, Argentina
\item[$^{13}$] University of Adelaide, Adelaide, S.A., Australia
\item[$^{14}$] Universit\'e Libre de Bruxelles (ULB), Brussels, Belgium
\item[$^{15}$] Vrije Universiteit Brussels, Brussels, Belgium
\item[$^{16}$] Centro Brasileiro de Pesquisas Fisicas, Rio de Janeiro, RJ, Brazil
\item[$^{17}$] Centro Federal de Educa\c{c}\~ao Tecnol\'ogica Celso Suckow da Fonseca, Petropolis, Brazil
\item[$^{18}$] Instituto Federal de Educa\c{c}\~ao, Ci\^encia e Tecnologia do Rio de Janeiro (IFRJ), Brazil
\item[$^{19}$] Universidade de S\~ao Paulo, Escola de Engenharia de Lorena, Lorena, SP, Brazil
\item[$^{20}$] Universidade de S\~ao Paulo, Instituto de F\'\i{}sica de S\~ao Carlos, S\~ao Carlos, SP, Brazil
\item[$^{21}$] Universidade de S\~ao Paulo, Instituto de F\'\i{}sica, S\~ao Paulo, SP, Brazil
\item[$^{22}$] Universidade Estadual de Campinas (UNICAMP), IFGW, Campinas, SP, Brazil
\item[$^{23}$] Universidade Estadual de Feira de Santana, Feira de Santana, Brazil
\item[$^{24}$] Universidade Federal de Campina Grande, Centro de Ciencias e Tecnologia, Campina Grande, Brazil
\item[$^{25}$] Universidade Federal do ABC, Santo Andr\'e, SP, Brazil
\item[$^{26}$] Universidade Federal do Paran\'a, Setor Palotina, Palotina, Brazil
\item[$^{27}$] Universidade Federal do Rio de Janeiro, Instituto de F\'\i{}sica, Rio de Janeiro, RJ, Brazil
\item[$^{28}$] Universidad de Medell\'\i{}n, Medell\'\i{}n, Colombia
\item[$^{29}$] Universidad Industrial de Santander, Bucaramanga, Colombia
\item[$^{30}$] Charles University, Faculty of Mathematics and Physics, Institute of Particle and Nuclear Physics, Prague, Czech Republic
\item[$^{31}$] Institute of Physics of the Czech Academy of Sciences, Prague, Czech Republic
\item[$^{32}$] Palacky University, Olomouc, Czech Republic
\item[$^{33}$] CNRS/IN2P3, IJCLab, Universit\'e Paris-Saclay, Orsay, France
\item[$^{34}$] Laboratoire de Physique Nucl\'eaire et de Hautes Energies (LPNHE), Sorbonne Universit\'e, Universit\'e de Paris, CNRS-IN2P3, Paris, France
\item[$^{35}$] Univ.\ Grenoble Alpes, CNRS, Grenoble Institute of Engineering Univ.\ Grenoble Alpes, LPSC-IN2P3, 38000 Grenoble, France
\item[$^{36}$] Universit\'e Paris-Saclay, CNRS/IN2P3, IJCLab, Orsay, France
\item[$^{37}$] Bergische Universit\"at Wuppertal, Department of Physics, Wuppertal, Germany
\item[$^{38}$] Karlsruhe Institute of Technology (KIT), Institute for Experimental Particle Physics, Karlsruhe, Germany
\item[$^{39}$] Karlsruhe Institute of Technology (KIT), Institut f\"ur Prozessdatenverarbeitung und Elektronik, Karlsruhe, Germany
\item[$^{40}$] Karlsruhe Institute of Technology (KIT), Institute for Astroparticle Physics, Karlsruhe, Germany
\item[$^{41}$] RWTH Aachen University, III.\ Physikalisches Institut A, Aachen, Germany
\item[$^{42}$] Universit\"at Hamburg, II.\ Institut f\"ur Theoretische Physik, Hamburg, Germany
\item[$^{43}$] Universit\"at Siegen, Department Physik -- Experimentelle Teilchenphysik, Siegen, Germany
\item[$^{44}$] Gran Sasso Science Institute, L'Aquila, Italy
\item[$^{45}$] INFN Laboratori Nazionali del Gran Sasso, Assergi (L'Aquila), Italy
\item[$^{46}$] INFN, Sezione di Catania, Catania, Italy
\item[$^{47}$] INFN, Sezione di Lecce, Lecce, Italy
\item[$^{48}$] INFN, Sezione di Milano, Milano, Italy
\item[$^{49}$] INFN, Sezione di Napoli, Napoli, Italy
\item[$^{50}$] INFN, Sezione di Roma ``Tor Vergata'', Roma, Italy
\item[$^{51}$] INFN, Sezione di Torino, Torino, Italy
\item[$^{52}$] Istituto di Astrofisica Spaziale e Fisica Cosmica di Palermo (INAF), Palermo, Italy
\item[$^{53}$] Osservatorio Astrofisico di Torino (INAF), Torino, Italy
\item[$^{54}$] Politecnico di Milano, Dipartimento di Scienze e Tecnologie Aerospaziali , Milano, Italy
\item[$^{55}$] Universit\`a del Salento, Dipartimento di Matematica e Fisica ``E.\ De Giorgi'', Lecce, Italy
\item[$^{56}$] Universit\`a dell'Aquila, Dipartimento di Scienze Fisiche e Chimiche, L'Aquila, Italy
\item[$^{57}$] Universit\`a di Catania, Dipartimento di Fisica e Astronomia ``Ettore Majorana``, Catania, Italy
\item[$^{58}$] Universit\`a di Milano, Dipartimento di Fisica, Milano, Italy
\item[$^{59}$] Universit\`a di Napoli ``Federico II'', Dipartimento di Fisica ``Ettore Pancini'', Napoli, Italy
\item[$^{60}$] Universit\`a di Palermo, Dipartimento di Fisica e Chimica ''E.\ Segr\`e'', Palermo, Italy
\item[$^{61}$] Universit\`a di Roma ``Tor Vergata'', Dipartimento di Fisica, Roma, Italy
\item[$^{62}$] Universit\`a Torino, Dipartimento di Fisica, Torino, Italy
\item[$^{63}$] Benem\'erita Universidad Aut\'onoma de Puebla, Puebla, M\'exico
\item[$^{64}$] Unidad Profesional Interdisciplinaria en Ingenier\'\i{}a y Tecnolog\'\i{}as Avanzadas del Instituto Polit\'ecnico Nacional (UPIITA-IPN), M\'exico, D.F., M\'exico
\item[$^{65}$] Universidad Aut\'onoma de Chiapas, Tuxtla Guti\'errez, Chiapas, M\'exico
\item[$^{66}$] Universidad Michoacana de San Nicol\'as de Hidalgo, Morelia, Michoac\'an, M\'exico
\item[$^{67}$] Universidad Nacional Aut\'onoma de M\'exico, M\'exico, D.F., M\'exico
\item[$^{68}$] Institute of Nuclear Physics PAN, Krakow, Poland
\item[$^{69}$] University of \L{}\'od\'z, Faculty of High-Energy Astrophysics,\L{}\'od\'z, Poland
\item[$^{70}$] Laborat\'orio de Instrumenta\c{c}\~ao e F\'\i{}sica Experimental de Part\'\i{}culas -- LIP and Instituto Superior T\'ecnico -- IST, Universidade de Lisboa -- UL, Lisboa, Portugal
\item[$^{71}$] ``Horia Hulubei'' National Institute for Physics and Nuclear Engineering, Bucharest-Magurele, Romania
\item[$^{72}$] Institute of Space Science, Bucharest-Magurele, Romania
\item[$^{73}$] Center for Astrophysics and Cosmology (CAC), University of Nova Gorica, Nova Gorica, Slovenia
\item[$^{74}$] Experimental Particle Physics Department, J.\ Stefan Institute, Ljubljana, Slovenia
\item[$^{75}$] Universidad de Granada and C.A.F.P.E., Granada, Spain
\item[$^{76}$] Instituto Galego de F\'\i{}sica de Altas Enerx\'\i{}as (IGFAE), Universidade de Santiago de Compostela, Santiago de Compostela, Spain
\item[$^{77}$] IMAPP, Radboud University Nijmegen, Nijmegen, The Netherlands
\item[$^{78}$] Nationaal Instituut voor Kernfysica en Hoge Energie Fysica (NIKHEF), Science Park, Amsterdam, The Netherlands
\item[$^{79}$] Stichting Astronomisch Onderzoek in Nederland (ASTRON), Dwingeloo, The Netherlands
\item[$^{80}$] Universiteit van Amsterdam, Faculty of Science, Amsterdam, The Netherlands
\item[$^{81}$] Case Western Reserve University, Cleveland, OH, USA
\item[$^{82}$] Colorado School of Mines, Golden, CO, USA
\item[$^{83}$] Department of Physics and Astronomy, Lehman College, City University of New York, Bronx, NY, USA
\item[$^{84}$] Michigan Technological University, Houghton, MI, USA
\item[$^{85}$] New York University, New York, NY, USA
\item[$^{86}$] University of Chicago, Enrico Fermi Institute, Chicago, IL, USA
\item[$^{87}$] University of Delaware, Department of Physics and Astronomy, Bartol Research Institute, Newark, DE, USA
\item[] -----
\item[$^{a}$] Max-Planck-Institut f\"ur Radioastronomie, Bonn, Germany
\item[$^{b}$] also at Kapteyn Institute, University of Groningen, Groningen, The Netherlands
\item[$^{c}$] School of Physics and Astronomy, University of Leeds, Leeds, United Kingdom
\item[$^{d}$] Fermi National Accelerator Laboratory, Fermilab, Batavia, IL, USA
\item[$^{e}$] Pennsylvania State University, University Park, PA, USA
\item[$^{f}$] Colorado State University, Fort Collins, CO, USA
\item[$^{g}$] Louisiana State University, Baton Rouge, LA, USA
\item[$^{h}$] now at Graduate School of Science, Osaka Metropolitan University, Osaka, Japan
\item[$^{i}$] Institut universitaire de France (IUF), France
\item[$^{j}$] now at Technische Universit\"at Dortmund and Ruhr-Universit\"at Bochum, Dortmund and Bochum, Germany
\end{description}

\section*{Acknowledgments}

\begin{sloppypar}
The successful installation, commissioning, and operation of the Pierre
Auger Observatory would not have been possible without the strong
commitment and effort from the technical and administrative staff in
Malarg\"ue. We are very grateful to the following agencies and
organizations for financial support:
\end{sloppypar}

\begin{sloppypar}
Argentina -- Comisi\'on Nacional de Energ\'\i{}a At\'omica; Agencia Nacional de
Promoci\'on Cient\'\i{}fica y Tecnol\'ogica (ANPCyT); Consejo Nacional de
Investigaciones Cient\'\i{}ficas y T\'ecnicas (CONICET); Gobierno de la
Provincia de Mendoza; Municipalidad de Malarg\"ue; NDM Holdings and Valle
Las Le\~nas; in gratitude for their continuing cooperation over land
access; Australia -- the Australian Research Council; Belgium -- Fonds
de la Recherche Scientifique (FNRS); Research Foundation Flanders (FWO),
Marie Curie Action of the European Union Grant No.~101107047; Brazil --
Conselho Nacional de Desenvolvimento Cient\'\i{}fico e Tecnol\'ogico (CNPq);
Financiadora de Estudos e Projetos (FINEP); Funda\c{c}\~ao de Amparo \`a
Pesquisa do Estado de Rio de Janeiro (FAPERJ); S\~ao Paulo Research
Foundation (FAPESP) Grants No.~2019/10151-2, No.~2010/07359-6 and
No.~1999/05404-3; Minist\'erio da Ci\^encia, Tecnologia, Inova\c{c}\~oes e
Comunica\c{c}\~oes (MCTIC); Czech Republic -- GACR 24-13049S, CAS LQ100102401,
MEYS LM2023032, CZ.02.1.01/0.0/0.0/16{\textunderscore}013/0001402,
CZ.02.1.01/0.0/0.0/18{\textunderscore}046/0016010 and
CZ.02.1.01/0.0/0.0/17{\textunderscore}049/0008422 and CZ.02.01.01/00/22{\textunderscore}008/0004632;
France -- Centre de Calcul IN2P3/CNRS; Centre National de la Recherche
Scientifique (CNRS); Conseil R\'egional Ile-de-France; D\'epartement
Physique Nucl\'eaire et Corpusculaire (PNC-IN2P3/CNRS); D\'epartement
Sciences de l'Univers (SDU-INSU/CNRS); Institut Lagrange de Paris (ILP)
Grant No.~LABEX ANR-10-LABX-63 within the Investissements d'Avenir
Programme Grant No.~ANR-11-IDEX-0004-02; Germany -- Bundesministerium
f\"ur Bildung und Forschung (BMBF); Deutsche Forschungsgemeinschaft (DFG);
Finanzministerium Baden-W\"urttemberg; Helmholtz Alliance for
Astroparticle Physics (HAP); Helmholtz-Gemeinschaft Deutscher
Forschungszentren (HGF); Ministerium f\"ur Kultur und Wissenschaft des
Landes Nordrhein-Westfalen; Ministerium f\"ur Wissenschaft, Forschung und
Kunst des Landes Baden-W\"urttemberg; Italy -- Istituto Nazionale di
Fisica Nucleare (INFN); Istituto Nazionale di Astrofisica (INAF);
Ministero dell'Universit\`a e della Ricerca (MUR); CETEMPS Center of
Excellence; Ministero degli Affari Esteri (MAE), ICSC Centro Nazionale
di Ricerca in High Performance Computing, Big Data and Quantum
Computing, funded by European Union NextGenerationEU, reference code
CN{\textunderscore}00000013; M\'exico -- Consejo Nacional de Ciencia y Tecnolog\'\i{}a
(CONACYT) No.~167733; Universidad Nacional Aut\'onoma de M\'exico (UNAM);
PAPIIT DGAPA-UNAM; The Netherlands -- Ministry of Education, Culture and
Science; Netherlands Organisation for Scientific Research (NWO); Dutch
national e-infrastructure with the support of SURF Cooperative; Poland
-- Ministry of Education and Science, grants No.~DIR/WK/2018/11 and
2022/WK/12; National Science Centre, grants No.~2016/22/M/ST9/00198,
2016/23/B/ST9/01635, 2020/39/B/ST9/01398, and 2022/45/B/ST9/02163;
Portugal -- Portuguese national funds and FEDER funds within Programa
Operacional Factores de Competitividade through Funda\c{c}\~ao para a Ci\^encia
e a Tecnologia (COMPETE); Romania -- Ministry of Research, Innovation
and Digitization, CNCS-UEFISCDI, contract no.~30N/2023 under Romanian
National Core Program LAPLAS VII, grant no.~PN 23 21 01 02 and project
number PN-III-P1-1.1-TE-2021-0924/TE57/2022, within PNCDI III; Slovenia
-- Slovenian Research Agency, grants P1-0031, P1-0385, I0-0033, N1-0111;
Spain -- Ministerio de Ciencia e Innovaci\'on/Agencia Estatal de
Investigaci\'on (PID2019-105544GB-I00, PID2022-140510NB-I00 and
RYC2019-027017-I), Xunta de Galicia (CIGUS Network of Research Centers,
Consolidaci\'on 2021 GRC GI-2033, ED431C-2021/22 and ED431F-2022/15),
Junta de Andaluc\'\i{}a (SOMM17/6104/UGR and P18-FR-4314), and the European Union (Marie Sklodowska-Curie 101065027 and ERDF); USA -- Department of Energy, Contracts No.~DE-AC02-07CH11359, No.~DE-FR02-04ER41300, No.~DE-FG02-99ER41107 and No.~DE-SC0011689; National Science Foundation, Grant No.~0450696, and NSF-2013199; The Grainger Foundation; Marie Curie-IRSES/EPLANET; European Particle Physics Latin American Network; and UNESCO.\\
The authors want to thank the lab. Atmósfera | DEILAP - CITEDEF that provided us the access to the GeoRayos database for consulting WWLLN, ENTLN and GLM data and Jeff Lapierre (Advanced Environmental Monitoring (AEM) - Germantown MD USA) for providing the ENTLN waveform in coincidence with the Auger event detected on December 8, 2025. 
\end{sloppypar}

}
%
%
%


\begin{thebibliography}{99}
\footnotesize
\raggedright
\setlength{\itemsep}{0pt}
\def\vyp#1#2#3{\textbf{#1} (#2) #3} 

\bibitem{BATSE}
G.~J.~Fishman~\textit{et al.}, Science \vyp{264}{1994}{5163}, 1313–1316.
\bibitem{PAO}
A.~Aab \textit{et al.}, Nucl.\ Instrum.\ Meth.\ A \vyp{798}{2015}{172–213}.
\bibitem{atmoHEAD24_moni}
M.~Büsken and B.~Keilhauer [for the Pierre Auger Coll.], J.\ Phys.\: Conf.\ Ser.\ \vyp{2985}{2025}{012001}.
\bibitem{ICRC2021}
R.~Colalillo [for the Pierre Auger Coll.], Proc. 37th Int. Cosmic Ray Conf., Berlin, Germany (2021), PoS(ICRC2021)395.
\bibitem{SDtrigger}
Pierre Auger Coll.,
\href{http://dx.doi.org/10.1016/j.nima.2009.11.018}{Nucl.\ Instrum.\ Meth.\ A \vyp{613}{2010}{29}}.
\bibitem{Dwyer21}
J.~Dwyer, Phys.\ Rev.\ D \textbf{104} (2021) 043012.
\bibitem{Stadnichuk21}
E.~Stadnichuk~\textit{et al.}, J.\ Geophys.\ Res.\ Atmospheres \textbf{126} (2021) e2021JD035278.
\bibitem{atmoHEAD22}
M.~Schimassek [for the Pierre Auger Coll.], J.\ Phys.\: Conf.\ Ser.\ \vyp{2398}{2022}{012003}.
\bibitem{ICRC2023}
R.~Colalillo [for the Pierre Auger Coll.], Proc. 38th Int. Cosmic Ray Conf., Nagoya, Japan (2023), PoS(ICRC2023)439.
\bibitem{WWLLN}
World Wide Lightning Location Network, https://wwlln.net/
\bibitem{ENTLN}
Earth Networks Total Lightning Network, https://aem.eco/product/earth-networks-total-lightning-network/
\bibitem{GLM}
Geostationary Lightning Mapper, GOES-16, https://www.earthdata.nasa.gov/data/instruments/glm
\bibitem{SmithAGU2024} 
T.~Wu~\textit{et al.}, Geophys.\ Ref.\ Lett. \textbf{52} (2025) e2024GL113194.
\bibitem{THOR}
D.~M.~Smith~\textit{et al.}, "New Behaviors in Downward Terrestrial Gamma-ray Flashes: Observations at Mt. Fuji and Uchinada, Japan ", AGU Fall Meeting 2023, AE22A-01
\bibitem{similTHOR}
J.~Dwyer~\textit{et al.}, J.\ Geophys.\ Res.\ \textbf{117} (2012) A10303.
\bibitem{Urbani2021-yj}
M.~Urbani~\textit{et al.}, J.\ Geophys.\ Res.\ \textbf{126} (2021) e2020JD033799.
\bibitem{Wada2025}
Y.~Wada~\textit{et al.}, Sci.\ Adv.\ \textbf{11} (2025) eads6906.
\bibitem{JamesAGU2024}
J.~Sanchez~\textit{et. al.}, "Probing the Context of TGF Events at the Pierre Auger Observatory Using VLF Sensors", AGU Fall Meeting 2024, AE23B-2585
\bibitem{atmoHEAD24}
M. Weitz [for the Pierre Auger Coll.], J.\ Phys.\: Conf.\ Ser.\ \vyp{2985}{2024}{012016}.
\bibitem{aera_beaconpaper}
A. Aab \textit{et al.} [Pierre Auger Coll.], J.\ Instrum.\ \textbf{11} (2016) P01018.
\bibitem{LDS_2015}
J. Rautenberg [for the Pierre Auger Coll.], Proc. 34th Int. Cosmic Ray Conf., Hague, Netherlands (2015), PoS(ICRC2015)678
\end{thebibliography}
\end{document}